\documentclass{llncs}

\usepackage{epsfig}
\usepackage{amsmath,amssymb}
\usepackage{subfigure}
\usepackage{framed}

\usepackage{url}
\usepackage{placeins}
\usepackage{tikz}
\usepackage{pgfplots}
\usetikzlibrary{snakes}

\newcommand{\G}{\mathcal{G}}

\date{\today}

\begin{document}

\title{Distributed Maintenance of Anytime Available Spanning Trees in Dynamic Networks}

\author{
    Arnaud Casteigts$^{1}$,~
    Serge Chaumette$^{1}$,~
    Fr\'ed\'eric Guinand$^{2}$~ and~
    Yoann Pign\'e$^{2}$
}
\institute{
    $^1$LaBRI, University of Bordeaux, France\\
    {\tt\{arnaud.casteigts,serge.chaumette\}@labri.fr}\medskip\\
    $^2$LITIS, University of Le Havre, France\\
    {\tt\{frederic.guinand,yoann.pigne\}@univ-lehavre.fr}\medskip\\
}


\maketitle

\begin{abstract}
  We address the problem of building and maintaining a forest of spanning trees in highly dynamic networks, in which topological events can occur at any time and any rate, and no stable periods can be assumed. In these harsh environments, we strive to preserve some properties such as cycle-freeness or existence of a unique root in each fragment regardless of the events, so as to keep these fragments functioning uninterruptedly to a possible extent. Our algorithm operates at a coarse-grain level, using atomic pairwise interactions akin to population protocol or graph relabeling systems. The algorithm relies on a perpetual alternation of \emph{topology-induced splittings} and \emph{computation-induced mergings} of a forest of trees. Each tree in the forest hosts exactly one token (also called root) that performs a random walk {\em inside} the tree, switching parent-child relationships as it crosses edges. When two tokens are located on both sides of a same edge, their trees are merged upon this edge and one token disappears. Whenever an edge that belongs to a tree disappears, its child endpoint regenerates a new token instantly. The main features of this approach is that both \emph{merging} and \emph{splitting} are purely localized phenomenons. This paper presents the algorithm and establishes its correctness in arbitrary dynamic networks. We also discuss aspects related to the implementation of this general principle in fine-grain models, as well as embryonic elements of analysis. The characterization of the algorithm performance is left open, both analytically and experimentally. 
\end{abstract}

\section{Introduction\label{sec:intro}}

Spanning trees are essential components in communication networks. The availability of such structures simplifies a large number of tasks, among which broadcasting, routing, or termination detection. From the standpoint of distributed computing, constructing a spanning tree implies the collaboration of neighboring nodes in order to establish selective relationships that inter-connect the whole network without cycle. 

The problem is very different in essence in static and dynamic networks. In a static network, there is generally a distinction between the construction of a tree and its effective use, both taking place at different times. In truly dynamic networks (e.g. vehicular networks), the set of communication links evolves rapidly and continuously. As a result, the trees need to be updated on a constant basis and while they are used. Early works addressing the spanning tree problem in dynamic graphs (see e.g.~\cite{time-optimal-1993,survey-gartner,time-optimal-2007} and the references therein) applied strong restrictions on the dynamicity; namely, these works assumed the network stabilizes eventually, or recurrently offers stable periods during which the tree can entirely be recomputed. These assumptions are certainly appropriate in the case of occasional failures or reconfigurations of the topology. But they are not reasonable in highly dynamic scenarios like mobile ad hoc networks.

We are interested in understanding what can still be done in the harshest dynamic context. In particular, we consider networks in which no stability period is ever expected; no information is available about future topological events; no restrictions apply to the rate of these events; and no contemporaneous end-to-end connectivity is assumed (that is, we address \emph{delay-tolerant networks}~\cite{Kevin-Fall}). On the other hand, we allow ourselves to reason at a high level of abstraction, using a coarse-grain interaction model akin to recent {\em population protocol} models~\cite{angluin2006}. While we find the problem in this model interesting in its own right, we still hope and believe the principles highlighted here can help subsequent effort to make it work in finer-grain (e.g. message passing) models.

The algorithm relies on a perpetual alternation of \emph{topology-induced splittings} and \emph{computation-induced mergings} of a forest of spanning trees. Each tree in the forest hosts exactly one token (also called root) that performs a random walk {\em inside} the tree, switching parent-child relationships as it crosses edges. When two tokens are located on both sides of a same edge, their trees are merged upon this edge and one token disappears. Whenever an edge that belongs to a tree disappears, its child endpoint regenerates a new token instantly. The main features of this approach is that both \emph{merging} and \emph{splitting} are purely localized phenomenons. 

After reviewing some relevant work in Section~\ref{sec:related}, we define the network model and assumptions, as well as the computational model in Section~\ref{sec:model}. The algorithm is then presented in detail and proved correct in Section~\ref{sec:original}. This presentation is followed by a discussion regarding some important implementation choices (e.g. priority between different rules of interaction). In Section~\ref{sec:analysis}, we provide preliminary results on the analysis of the algorithm, which we regard as a coalescing particle system involving random walks in trees.
We conclude in Section~\ref{sec:conclusion} with some perspectives. 

\section{Related Work\label{sec:related}}
The problem of building distributed spanning trees in communication networks, and more generally in graphs, has been extensively studied during the last three decades and a large literature exists on the topic. It is noteworthy that the problem was studied by different communities (self-stabilization, stochastic processes, distributed computing) using different paradigms and terminologies (\textit{e.g.} \emph{token, mobile agent, random walk, legal state, stabilization time, coalescing time, tree, forest, etc.}). We review below the most relevant concepts and approaches to solve this problem.

\paragraph{Self-stabilization:}
A system that reaches a \emph{legal} state starting from an \emph{arbitrary} state is called \emph{self-stabilizing}. After a fault in the system, the time required to reach the legal state is called the \emph{stabilization time}. In the context of spanning trees in dynamic networks, topological changes are the faults, and having the entire network covered by a single tree, or in case of partitioned networks one tree per connected component, is the legal state. One approach to transform a non-self-stabilizing algorithm into a self-stabilizing one, is to \emph{reset} the states of the nodes when a fault occurs, so that a new execution of the algorithm is initiated. This approach has been considered by most self-stabilizing algorithms proposed so far for the spanning tree problem, and an optimal-time solution was proposed in~\cite{time-optimal-1993} (as a coarse-grain graph algorithm, more recently transposed into the message passing model in~\cite{time-optimal-2007}). We refer the reader to \cite{survey-gartner} for a more general survey on self-stabilizing spanning tree algorithms. In these works, the algorithms assume that no additional fault occur during the stabilization period, which is not acceptable in highly dynamic networks. 

\paragraph{Random walk:}
A random walk is a sequence of nodes such that each node in the sequence (except the starting node) is randomly selected among the neighbors of its predecessor. Random walks have been used to solve several problems in distributed systems, such as leader election, voting, or spanning trees~\cite{cooper2012}. The idea of using random walks to compute spanning trees was first proposed by Aldous in~\cite{random-walk-trees}, where a single random walk is considered. Anytime, the set of all covered nodes, along with the edges from which they were visited the first time, defines a random tree that spans the nodes already visited.

\paragraph{Mobile agents:}
Mobile agents are entities that can travel across the network, and perform tasks on the underlying nodes. These agents may or may not carry their own memory, and adopt a variety of strategies to move within the network. In~\cite{Baala2003}, distributed random walks of mobile agents (called \emph{tokens} in the paper) are used. More precisely, colored tokens are annexing territories while walking within the network. Each token builds a tree (a subtree of the global spanning tree). When two tokens meet or when a token visits a node that have already been visited, the two trees are merged into one. This operation is performed by a \emph{wave propagation}, which is a broadcast-based process that occurs along the edges of the trees. The network is assumed connected and no topological changes are allowed during the construction of the tree. Unique identifiers are also required. 
A related approach was proposed in~\cite{mosbah-tree}, where mobile colored agents (equivalent to tokens) construct subtrees that are progressively merged into a final spanning tree. Whenever one agent enters the region of another, the agent that have the larger color progressively takes control of the nodes and eventually destroys the other agent. The advantage of this gradual process is that it avoids the wave propagation. However, unique identifiers are still required to generate the colors and some global information (an upper bound in the cover time of the random walk) is needed to regenerate an agent. Finally, the approach does not tolerate frequent topological events.

In comparison to these approaches, the one we propose does not require stable periods or unique identifiers (nor any global information). This is, to the best of our knowledge, the first attempt in this direction.

\section{Network model and assumptions\label{sec:model}}

We represent the network as an evolving graph $\G=\{G_1,G_2,...\}$, all elements of which correspond to snapshots of the topology, and the transitions between them bijectively reflect the occurrence of one, or several simultaneous topological events (appearance or disappearance of edges). More elaborate variants of evolving graphs can be found in the original paper~\cite{Fer04}. However, this basic variant is suitable enough for our purpose.

At a given moment, the network is therefore represented by an undirected simple graph $G_i=(V,E_i)$, where the set of nodes $V$ is assumed to be constant, while the set of edges varies without restriction from one $G_i$ to the next. The temporal span of each $G_i$ is arbitrary and in particular, it is not bounded (whether from above or below). We do not require the existence of unique identifiers for the nodes, but we assume they are able to distinguish between their incident edges and assign a local value to them (thus, an edge typically has two values, one on each side). Note that in practice, especially in a wireless network, this feature would require unique identifiers to be implemented. It is however a weaker assumption from a theoretical standpoint. Further, it is more natural to think of our algorithm without identifiers.

\subsection{Computational model}
We consider a coarse-grain interaction model akin to {\em population protocols}~\cite{angluin2006} or {\em graph relabeling systems}~\cite{LMS99}. In these models a computation step is an atomic pairwise interaction. Precisely, a computation step takes as input the state of a pair of nodes (together with their common edge), and modifies these states according to some rule. For example, the rule 
\begin{minipage}[c]{6.5cm}
  \begin{tikzpicture}[scale=1]
  \tikzstyle{every node}=[draw,circle,fill=black!80,inner sep=1.2pt]
  \path (-1.5,0) node (v11) {};
  \path (-.2,0) node (v12) {};
  \path (1.7,0) node (v21) {};
  \path (3,0) node (v22) {};
  \draw[thick, ->, shorten >=15pt, shorten <=15pt] (v12)--(v21);
  \draw (v11)--(v12);
  \draw (v21)--(v22);
  \tikzstyle{every node}=[font=\scriptsize]
  \path (v11.north)+(0,.12) node (d) {\texttt{inside}};
  \path (v12.north)+(0,.12) node (d) {\texttt{outside}};
  \path (v21.north)+(0,.12) node (d) {\texttt{inside}};
  \path (v22.north)+(0,.12) node (d) {\texttt{inside}};
  \tikzstyle{every node}=[font=\tiny]
  \path (v11)+(.12,-.1) node {$0$};
  \path (v12)+(-.12,-.1) node {$0$};
  \path (v21)+(.12,-.1) node {$2$};
  \path (v22)+(-.12,-.1) node {$1$};
\end{tikzpicture}
\end{minipage}
may represent the construction of a rooted spanning tree in a static network from some distinguished {\small \tt inside} node. We assume in general that two interactions can occur in parallel so long as they are disjoint (they do not imply a common node). The way interactions are selected, that is, the {\em scheduling}, is typically not a part of the algorithm (e.g. it can be adversarial with some constraints, or probabilistic, or result from some finer-grain interaction). The general properties we establish on our algorithm are insensitive to these concerns. Note that the guard of a rule (left part) may represent two nodes in a same state. In this case, despite the absence of unique identifiers, symmetry is broken by the application of the rule -- however, the choice of what role is played by each node is not controlled by the algorithm (it is up to the scheduler).

Dealing with a dynamic graph (the usual population protocols deal with static graphs), we consider another type of operation in addition to pairwise interaction. This operation, triggered by topological events, consists in updating the state of a node immediately after one of its edges disappears. As such, an algorithm can associate reactive operations to the loss of a link.

\section{The spanning forest algorithm}
\label{sec:original}

Informally, the algorithm is based on three operations on tokens: \textit{circulation}, \textit{merging}, and \textit{regeneration}, which aim at maintaining exactly one token per tree. 
Initially, every node forms a tree of its own and is the root of that tree (it has the token). When two token owners interact over a common edge, their tokens are merged into one and their common edge is added to the tree ({\em merging} rule $r_1$, see Figure~\ref{fig:merging} below). The parent-child relation is set accordingly. The rest of the time, each token performs a random walk along the edges of its own tree ({\em circulation} rule $r_2$, see Figure~\ref{fig:circulation} below) in search of new merging opportunities; parent-child relations are flipped as the circulation proceeds, so that a node can always tell, locally, which edge leads to the token. Whenever an edge of the tree disappears, the node on the child side regenerates a token ({\em regeneration} rule $r_a$, see Figure~\ref{fig:regeneration} below), which re-enables its orphan tree to keep running the process.

\subsection{State space and initialization}

At any time, the state of the system is fully described by two functions: one function for the state of the nodes $\lambda:V \rightarrow \{T,N\}$, where $T$ means this node has a token, while $N$ means it does not; and one function for the state of the edges {\em locally to both endpoints} $\lambda:V\times E_i \rightarrow \{0,1,2\}$, where $E_i$ is the current set of edges. The domain of both functions being different and non-ambiguous from the context, we authorize a unique symbol $\lambda$ to denote them. State $0$ for an edge means it does not belong to a tree. States $1$ or $2$ mean it does, and the local direction is from child to parent (state $1$) or from parent to child (state $2$). Hence, an edge whose state is $1$ at one end, must be in state $2$ at the other end. Notice that one bit of information is enough to encode the state of a node, and two bits, locally at each node, are sufficient for an edge.

\paragraph{Initialization:}
Given the first graph $G_0=(V,E_0)$, we set $\lambda(v)=T$ for all $v\in V$. We also set $\lambda(v,e)=0$ and $\lambda(u,e)=0$ for all $e=(u,v)\in E_0$. In words, every node initially holds a token and none of the edges belong to a tree.

\subsection{State transitions}

The evolution of the process is determined by two sources of events: topological events (i.e., appearance or disappearance of an edge) and computational events (i.e., pairwise interaction). We specify both separately. Keep in mind the principle presented here is intended to be extremely general, and several important questions, like priority among rules or the role played by each node in the rule, are deliberately set aside at this point. (They are discussed shortly after.)

\subsubsection{Transitions induced by pairwise interaction}\vspace{-3pt}

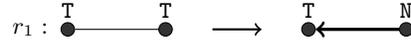
\begin{figure}[h]
  \vspace{-7pt}
\centering
\begin{tikzpicture}[scale=1]
  \tikzstyle{every node}=[draw,circle,fill=black!80,inner sep=1.8pt]
  \path (-1.5,0) node (v11) {};
  \path (-.2,0) node (v12) {};
  \path (1.7,0) node (v21) {};
  \path (3,0) node (v22) {};
  \draw[thick, ->, shorten >=15pt, shorten <=15pt] (v12)--(v21);
  \draw (v11)--(v12);
  \draw[very thick,<-] (v21)--(v22);
  \tikzstyle{every node}=[font=\footnotesize]
  \path (-2,0) node (rb) {$r_1:$ };
  \path (v11.north)+(0,.15) node (lab) {\texttt{T}};
  \path (v12.north)+(0,.15) node (lab) {\texttt{T}};
  \path (v21.north)+(0,.15) node (lab) {\texttt{T}};
  \path (v22.north)+(0,.15) node (lab) {\texttt{N}};
\end{tikzpicture}\vspace{-7pt}
\caption{\label{fig:merging}Merging rule (graphical representation).}
\end{figure}

\paragraph{Merging rule:}
Given two nodes $u$ and $v$ involved in an interaction over an edge $e=(u,v)$, the operation is specified as follows. If $\lambda(u)=T$ and $\lambda(v)=T$, then set $\lambda(v)=N$, $\lambda(v,e)=1$, and $\lambda(u,e)=2$. This rule, called {\em merging} rule ($r_1$), can be represented graphically as shown in Figure~\ref{fig:merging}.

\vspace{-3pt}

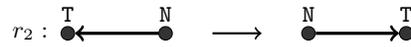
\begin{figure}[h]
  \vspace{-7pt}
\centering
\begin{tikzpicture}[scale=1]
  \tikzstyle{every node}=[draw,circle,fill=black!80,inner sep=1.8pt]
  \path (-1.5,0) node (v11) {};
  \path (-.2,0) node (v12) {};
  \path (1.7,0) node (v21) {};
  \path (3,0) node (v22) {};
  \draw[thick, ->, shorten >=15pt, shorten <=15pt] (v12)--(v21);
  \draw[very thick,<-] (v11)--(v12);
  \draw[very thick,->] (v21)--(v22);
  \tikzstyle{every node}=[font=\footnotesize]
  \path (-2,0) node (rb) {$r_2:$ };
  \path (v11.north)+(0,.15) node (lab) {\texttt{T}};
  \path (v12.north)+(0,.15) node (lab) {\texttt{N}};
  \path (v21.north)+(0,.15) node (lab) {\texttt{N}};
  \path (v22.north)+(0,.15) node (lab) {\texttt{T}};
\end{tikzpicture}\vspace{-7pt}
\caption{\label{fig:circulation}Circulation rule (graphical representation).}
\end{figure}

\paragraph{Circulation rule:}
Given two nodes $u$ and $v$ involved in an interaction over an edge $e=(u,v)$, the operation is specified as follows. If $\lambda(u)=T$ and $\lambda(v)=N$ and $\lambda(u,e)=2$, then set $\lambda(u)=N$, $\lambda(v)=T$, $\lambda(v,e)=2$, and $\lambda(u,e)=1$. This {\em circulation} rule ($r_2$) can be represented graphically as shown on Figure~\ref{fig:circulation}.

\subsubsection{Transitions induced by topological events}~\medskip

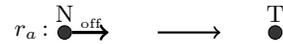
\begin{figure}[h]
  \vspace{-7pt}
    \centering
    \begin{tikzpicture}[scale=1]
      \tikzstyle{every node}=[draw,circle,fill=black!80,inner sep=1.8pt]
      \path (-1.5,0) node (v11) {};
      \path (1.3,0) node (v21) {};
      \tikzstyle{every node}=[]
      \path (-.8,0) node (v12) {};
      \draw[thick, ->, shorten >=18pt, shorten <=12pt] (v12)--(v21);
      \draw[very thick, ->] (v11)--(v12);
      \tikzstyle{every node}=[font=\footnotesize]
      \path (-1.6,0) node[left] (rb) {$r_a$\,:};
      \path (v11.north)+(0,.15) node (lab) {N};
      \path (v21.north)+(0,.15) node (lab) {T};
      \tikzstyle{every node}=[font=\tiny]
      \path (v11.east)+(.25,.12) node (lab) {off};
    \end{tikzpicture}\vspace{-7pt}
    \caption{\label{fig:regeneration} Regeneration rule (graphical representation).}
    \end{figure}

\noindent Given two consecutive graphs $G_i$ and $G_{i+1}$ in $\G$, the transition from one to the other induces the following updates on the states of the system. 
\paragraph{Appearance of an edge:} For all $e=(u,v) \in E_{i+1}$$\setminus$$E_{i}$, both $\lambda(u,e)$ and $\lambda(v,e)$ are set to $0$. In words, new edges are initialized with state $0$ on both sides.
\paragraph{Disappearance of an edge:} For all $e=(u,v) \in E_{i}$$\setminus$$E_{i+1}$, if $\lambda(u,e)=1$, then set $\lambda(u)=T$; else if $\lambda(v,e)=1$, then set $\lambda(v)=T$. In words, if a node loses the edge leading to its parent, it regenerates a token immediately. This rule, called {\em regeneration} rule ($r_a$), can be represented graphically as shown on Figure~\ref{fig:regeneration}.

An example execution sequence of the algorithm is provided on Figure~\ref{fig:sequence}.

\tikzset{every node/.style={draw, circle, fill, black!80, inner sep=0, minimum size=3pt}}
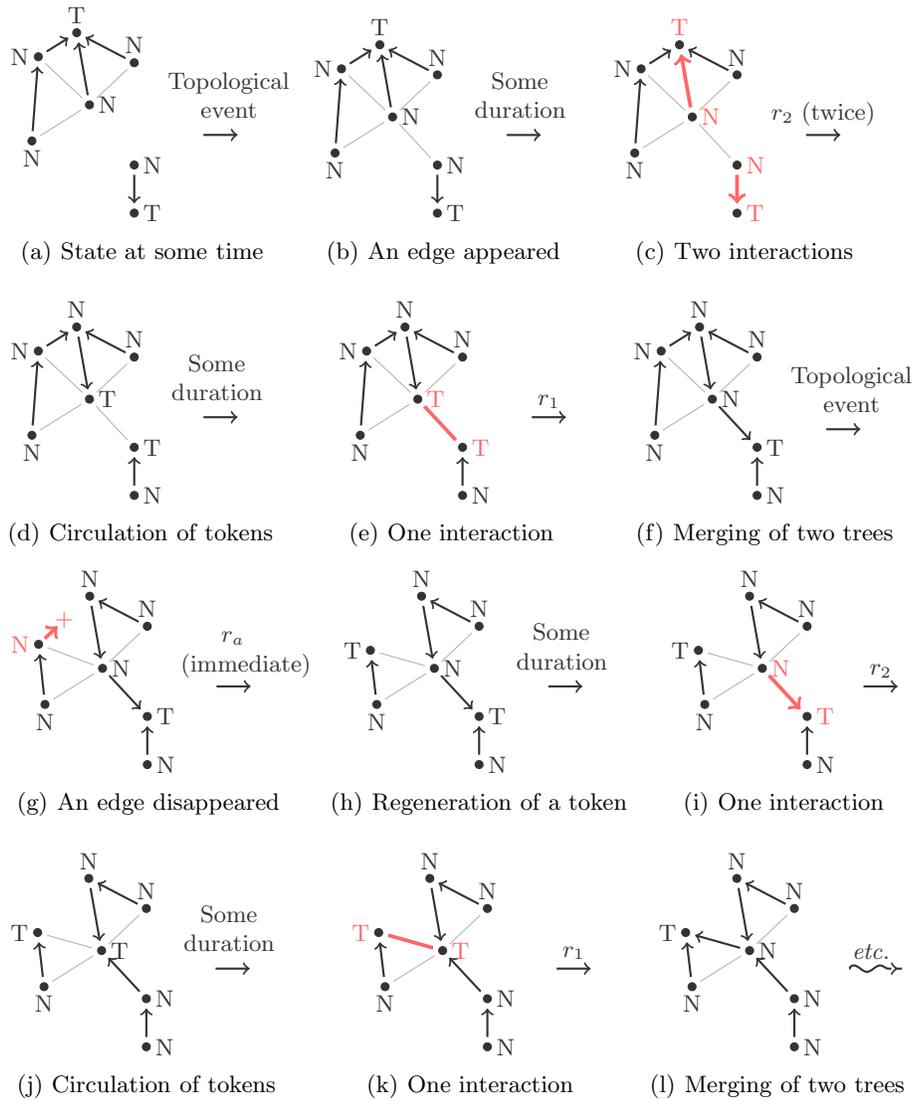
\begin{figure}[H]
  \subfigure[State at some time]{\label{fig:dspan1}
    \begin{tikzpicture}[xscale=0.85,yscale=0.8]
      \path (0,0) node (e){};
      \path (e)+(-.9,-.6) node (a){};
      \path (e)+(-.8,.8) node (b){};
      \path (e)+(-.2,1.2) node (c){};
      \path (e)+(.7,.7) node (d){};
      \path (e)+(.7,-1) node (f){};
      \path (f)+(0,-.8) node (g){};

      \tikzstyle{every node}=[font=\footnotesize]
      \tikzstyle{every path}=[gray!60,shorten >=2pt,shorten <=2pt]
      \draw (a)--(e);
      \draw (b)--(e);
      \draw (e)--(d);
      \tikzstyle{every path}=[thick,black!80,shorten >=2pt,shorten <=2pt]
      \draw[->] (a)--(b);
      \draw[->] (b)--(c);
      \draw[<-] (c)--(e);
      \draw[<-] (c)--(d);
      \draw[->] (f)--(g);
      \path (a) node[below] {N};
      \path (b) node[left] {N};
      \path (c) node[above] {T};
      \path (d) node[above] {N};
      \path (e) node[right] {N};
      \path (f) node[right] {N};
      \path (g) node[right] {T};

      \draw (1.7,-.5) edge[->] node[above=3pt, text width=1.4cm]{Topological\\~~~~event}(2.4,-.5);
    \end{tikzpicture}
  }
  \subfigure[An edge appeared]{\label{fig:dspan2}
    \begin{tikzpicture}[xscale=0.85,yscale=0.8]
      \path (0,0) node (e){};
      \path (e)+(-.9,-.6) node (a){};
      \path (e)+(-.8,.8) node (b){};
      \path (e)+(-.2,1.2) node (c){};
      \path (e)+(.7,.7) node (d){};
      \path (e)+(.7,-.8) node (f){};
      \path (f)+(0,-.8) node (g){};

      \tikzstyle{every node}=[font=\footnotesize]
      \tikzstyle{every path}=[gray!60,shorten >=2pt,shorten <=2pt]
      \draw (a)--(e);
      \draw (b)--(e);
      \draw (e)--(d);
      \draw (e)--(f);
      \tikzstyle{every path}=[thick,black!80,shorten >=2pt,shorten <=2pt]
      \draw[->] (a)--(b);
      \draw[->] (b)--(c);
      \draw[<-] (c)--(e);
      \draw[<-] (c)--(d);
      \draw[->] (f)--(g);
      \path (a) node[below] {N};
      \path (b) node[left] {N};
      \path (c) node[above] {T};
      \path (d) node[above] {N};
      \path (e) node[right] {N};
      \path (f) node[right] {N};
      \path (g) node[right] {T};

      \draw (1.7,-.3) edge[->] node[above=3pt, text width=1.3cm]{~~Some\\duration} (2.4,-.3);
    \end{tikzpicture}
  }
  \subfigure[Two interactions]{\label{fig:dspan3}
    \begin{tikzpicture}[xscale=0.85,yscale=0.8]
      \path (0,0) node (e){};
      \path (e)+(-.9,-.6) node (a){};
      \path (e)+(-.8,.8) node (b){};
      \path (e)+(-.2,1.2) node (c){};
      \path (e)+(.7,.7) node (d){};
      \path (e)+(.7,-.8) node (f){};
      \path (f)+(0,-.8) node (g){};

      \tikzstyle{every node}=[font=\footnotesize]
      \tikzstyle{every path}=[gray!60,shorten >=2pt,shorten <=2pt]
      \draw (a)--(e);
      \draw (b)--(e);
      \draw (e)--(d);
      \draw (e)--(f);
      \tikzstyle{every path}=[thick,black!80,shorten >=2pt,shorten <=2pt]
      \draw[->] (a)--(b);
      \draw[->] (b)--(c);
      \draw[<-,line width=1.4pt,red!60] (c)--(e);
      \draw[<-] (c)--(d);
      \draw[->,line width=1.4pt,red!60] (f)--(g);
      \path (a) node[below] {N};
      \path (b) node[left] {N};
      \path (c) node[above,red!60] {T};
      \path (d) node[above] {N};
      \path (e) node[right,red!60] {N};
      \path (f) node[right,red!60] {N};
      \path (g) node[right,red!60] {T};

      \draw (1.7,-.3) edge[->] node[above]{$r_2$ (twice)}(2.4,-.3);
    \end{tikzpicture}
  }\\
  \subfigure[Circulation of tokens]{\label{fig:dspan4}
    \begin{tikzpicture}[xscale=0.85,yscale=0.8]
      \path (0,0) node (e){};
      \path (e)+(-.9,-.6) node (a){};
      \path (e)+(-.8,.8) node (b){};
      \path (e)+(-.2,1.2) node (c){};
      \path (e)+(.7,.7) node (d){};
      \path (e)+(.7,-.8) node (f){};
      \path (f)+(0,-.8) node (g){};

      \tikzstyle{every node}=[font=\footnotesize]
      \tikzstyle{every path}=[gray!60,shorten >=2pt,shorten <=2pt]
      \draw (a)--(e);
      \draw (b)--(e);
      \draw (e)--(d);
      \draw (e)--(f);
      \tikzstyle{every path}=[thick,black!80,shorten >=2pt,shorten <=2pt]
      \draw[->] (a)--(b);
      \draw[->] (b)--(c);
      \draw[->] (c)--(e);
      \draw[<-] (c)--(d);
      \draw[<-] (f)--(g);
      \path (a) node[below] {N};
      \path (b) node[left] {N};
      \path (c) node[above] {N};
      \path (d) node[above] {N};
      \path (e) node[right] {T};
      \path (f) node[right] {T};
      \path (g) node[right] {N};

      \draw (1.7,-.3) edge[->] node[above=3pt, text width=1.3cm]{~~Some\\duration}(2.4,-.3);
    \end{tikzpicture}
  }
  \subfigure[One interaction]{\label{fig:dspan5}
    \begin{tikzpicture}[xscale=0.85,yscale=0.8]
      \path (0,0) node (e){};
      \path (e)+(-.9,-.6) node (a){};
      \path (e)+(-.8,.8) node (b){};
      \path (e)+(-.2,1.2) node (c){};
      \path (e)+(.7,.7) node (d){};
      \path (e)+(.7,-.8) node (f){};
      \path (f)+(0,-.8) node (g){};

      \tikzstyle{every node}=[font=\footnotesize]
      \tikzstyle{every path}=[gray!60,shorten >=2pt,shorten <=2pt]
      \draw (a)--(e);
      \draw (b)--(e);
      \draw (e)--(d);
      \draw[line width=1.4pt,red!60] (e)--(f);
      \tikzstyle{every path}=[thick,black!80,shorten >=2pt,shorten <=2pt]
      \draw[->] (a)--(b);
      \draw[->] (b)--(c);
      \draw[->] (c)--(e);
      \draw[<-] (c)--(d);
      \draw[<-] (f)--(g);
      \path (a) node[below] {N};
      \path (b) node[left] {N};
      \path (c) node[above] {N};
      \path (d) node[above] {N};
      \path (e) node[right,red!60] {T};
      \path (f) node[right,red!60] {T};
      \path (g) node[right] {N};

      \draw (1.7,-.3) edge[->] node[above]{$r_1$}(2.4,-.3);
    \end{tikzpicture}
  }
  \subfigure[Merging of two trees]{\label{fig:dspan6}
    \begin{tikzpicture}[xscale=0.85,yscale=0.8]
      \path (0,0) node (e){};
      \path (e)+(-.9,-.6) node (a){};
      \path (e)+(-.8,.8) node (b){};
      \path (e)+(-.2,1.2) node (c){};
      \path (e)+(.7,.7) node (d){};
      \path (e)+(.7,-.8) node (f){};
      \path (f)+(0,-.8) node (g){};

      \tikzstyle{every node}=[font=\footnotesize]
      \tikzstyle{every path}=[gray!60,shorten >=2pt,shorten <=2pt]
      \draw (a)--(e);
      \draw (b)--(e);
      \draw (e)--(d);
      \tikzstyle{every path}=[thick,black!80,shorten >=2pt,shorten <=2pt]
      \draw[->] (e)--(f);
      \draw[->] (a)--(b);
      \draw[->] (b)--(c);
      \draw[->] (c)--(e);
      \draw[<-] (c)--(d);
      \draw[<-] (f)--(g);
      \path (a) node[below] {N};
      \path (b) node[left] {N};
      \path (c) node[above] {N};
      \path (d) node[above] {N};
      \path (e) node[right] {N};
      \path (f) node[right] {T};
      \path (g) node[right] {N};

      \draw (1.7,-.5) edge[->] node[above=3pt, text width=1.4cm]{Topological\\~~~~event}(2.4,-.5);
    \end{tikzpicture}
  }
  \subfigure[An edge disappeared]{\label{fig:dspan7}
    \begin{tikzpicture}[xscale=0.85,yscale=0.8]
      \path (0,0) node (e){};
      \path (e)+(-.9,-.6) node (a){};
      \path (e)+(-1,.4) node (b){};
      \path (e)+(-.2,1.2) node (c){};
      \path (e)+(.7,.7) node (d){};
      \path (e)+(.7,-.8) node (f){};
      \path (f)+(0,-.8) node (g){};

      \tikzstyle{every node}=[font=\footnotesize]
      \tikzstyle{every path}=[gray!60,shorten >=2pt,shorten <=2pt]
      \draw (a)--(e);
      \draw (b)--(e);
      \draw (e)--(d);
      \tikzstyle{every path}=[thick,black!80,shorten >=2pt,shorten <=2pt]
      \draw[->] (e)--(f);
      \draw[->] (a)--(b);
      \draw[->,line width=1.4pt,red!60,shorten >=15pt,shorten <=1.5pt] (b)--node[pos=.5]{$+$}(c);
      \draw[->] (c)--(e);
      \draw[<-] (c)--(d);
      \draw[<-] (f)--(g);
      \path (a) node[below] {N};
      \path (b) node[left,red!60] {N};
      \path (c) node[above] {N};
      \path (d) node[above] {N};
      \path (e) node[right] {N};
      \path (f) node[right] {T};
      \path (g) node[right] {N};

      \draw (1.7,-.3) edge[->] node[above,text width=1.4cm]{~~~~~$r_a$\\(immediate)}(2.4,-.3);
    \end{tikzpicture}
  }
  \subfigure[Regeneration of a token]{\label{fig:dspan8}
    \begin{tikzpicture}[xscale=0.85,yscale=0.8]
      \path (0,0) node (e){};
      \path (e)+(-.9,-.6) node (a){};
      \path (e)+(-1,.3) node (b){};
      \path (e)+(-.2,1.2) node (c){};
      \path (e)+(.7,.7) node (d){};
      \path (e)+(.7,-.8) node (f){};
      \path (f)+(0,-.8) node (g){};

      \tikzstyle{every node}=[font=\footnotesize]
      \tikzstyle{every path}=[gray!60,shorten >=2pt,shorten <=2pt]
      \draw (a)--(e);
      \draw (b)--(e);
      \draw (e)--(d);
      \tikzstyle{every path}=[thick,black!80,shorten >=2pt,shorten <=2pt]
      \draw[->] (a)--(b);
      \draw[->] (c)--(e);
      \draw[<-] (c)--(d);
      \draw[->] (e)--(f);
      \draw[<-] (f)--(g);
      \path (a) node[below] {N};
      \path (b) node[left] {T};
      \path (c) node[above] {N};
      \path (d) node[above] {N};
      \path (e) node[right] {N};
      \path (f) node[right] {T};
      \path (g) node[right] {N};

      \draw (1.7,-.3) edge[->] node[above=3pt, text width=1.3cm]{~~Some\\duration} (2.4,-.3);
    \end{tikzpicture}
  }
  \subfigure[One interaction]{\label{fig:dspan9}
    \begin{tikzpicture}[xscale=0.85,yscale=0.8]
      \path (0,0) node (e){};
      \path (e)+(-.9,-.6) node (a){};
      \path (e)+(-1,.3) node (b){};
      \path (e)+(-.2,1.2) node (c){};
      \path (e)+(.7,.7) node (d){};
      \path (e)+(.7,-.8) node (f){};
      \path (f)+(0,-.8) node (g){};

      \tikzstyle{every node}=[font=\footnotesize]
      \tikzstyle{every path}=[gray!60,shorten >=2pt,shorten <=2pt]
      \draw (a)--(e);
      \draw (b)--(e);
      \draw (e)--(d);
      \tikzstyle{every path}=[thick,black!80,shorten >=2pt,shorten <=2pt]
      \draw[->] (a)--(b);
      \draw[->] (c)--(e);
      \draw[<-] (c)--(d);
      \draw[->,line width=1.4pt,red!60] (e)--(f);
      \draw[<-] (f)--(g);
      \path (a) node[below] {N};
      \path (b) node[left] {T};
      \path (c) node[above] {N};
      \path (d) node[above] {N};
      \path (e) node[right,red!60] {N};
      \path (f) node[right,red!60] {T};
      \path (g) node[right] {N};

      \draw (1.5,-.3) edge[->] node[above]{$r_2$}(2.2,-.3);
    \end{tikzpicture}
  }
  \subfigure[Circulation of tokens]{\label{fig:dspan3}
    \begin{tikzpicture}[xscale=0.85,yscale=0.8]
      \path (0,0) node (e){};
      \path (e)+(-.9,-.6) node (a){};
      \path (e)+(-1,.3) node (b){};
      \path (e)+(-.2,1.2) node (c){};
      \path (e)+(.7,.7) node (d){};
      \path (e)+(.7,-.8) node (f){};
      \path (f)+(0,-.8) node (g){};

      \tikzstyle{every node}=[font=\footnotesize]
      \tikzstyle{every path}=[gray!60,shorten >=2pt,shorten <=2pt]
      \draw (a)--(e);
      \draw (b)--(e);
      \draw (e)--(d);
      \tikzstyle{every path}=[thick,black!80,shorten >=2pt,shorten <=2pt]
      \draw[->] (a)--(b);
      \draw[->] (c)--(e);
      \draw[<-] (c)--(d);
      \draw[<-] (e)--(f);
      \draw[<-] (f)--(g);
      \path (a) node[below] {N};
      \path (b) node[left] {T};
      \path (c) node[above] {N};
      \path (d) node[above] {N};
      \path (e) node[right] {T};
      \path (f) node[right] {N};
      \path (g) node[right] {N};

      \draw (1.7,-.3) edge[->] node[above=3pt, text width=1.3cm]{~~Some\\duration} (2.4,-.3);
    \end{tikzpicture}
  }
  \hfill
  \subfigure[One interaction]{\label{fig:dspan4}
    \begin{tikzpicture}[xscale=0.85,yscale=0.8]
      \path (0,0) node (e){};
      \path (e)+(-.9,-.6) node (a){};
      \path (e)+(-1,.3) node (b){};
      \path (e)+(-.2,1.2) node (c){};
      \path (e)+(.7,.7) node (d){};
      \path (e)+(.7,-.8) node (f){};
      \path (f)+(0,-.8) node (g){};

      \tikzstyle{every node}=[font=\footnotesize]
      \tikzstyle{every path}=[gray!60,shorten >=2pt,shorten <=2pt]
      \draw (a)--(e);
      \draw[line width=1.4pt,red!60] (b)--(e);
      \draw (e)--(d);
      \tikzstyle{every path}=[thick,black!80,shorten >=2pt,shorten <=2pt]
      \draw[->] (a)--(b);
      \draw[->] (c)--(e);
      \draw[<-] (c)--(d);
      \draw[<-] (e)--(f);
      \draw[<-] (f)--(g);
      \path (a) node[below] {N};
      \path (b) node[left,red!60] {T};
      \path (c) node[above] {N};
      \path (d) node[above] {N};
      \path (e) node[right,red!60] {T};
      \path (f) node[right] {N};
      \path (g) node[right] {N};

      \draw (1.7,-.3) edge[->] node[above]{$r_1$}(2.4,-.3);
    \end{tikzpicture}
  }
  \hfill
  \subfigure[Merging of two trees]{\label{fig:dspan4}
    \begin{tikzpicture}[xscale=0.85,yscale=0.8]
      \path (0,0) node (e){};
      \path (e)+(-.9,-.6) node (a){};
      \path (e)+(-1,.3) node (b){};
      \path (e)+(-.2,1.2) node (c){};
      \path (e)+(.7,.7) node (d){};
      \path (e)+(.7,-.8) node (f){};
      \path (f)+(0,-.8) node (g){};

      \tikzstyle{every node}=[font=\footnotesize]
      \tikzstyle{every path}=[gray!60,shorten >=2pt,shorten <=2pt]
      \draw (a)--(e);
      \draw (e)--(d);
      \tikzstyle{every path}=[thick,black!80,shorten >=2pt,shorten <=2pt]
      \draw[<-] (b)--(e);
      \draw[->] (a)--(b);
      \draw[->] (c)--(e);
      \draw[<-] (c)--(d);
      \draw[<-] (e)--(f);
      \draw[<-] (f)--(g);
      \path (a) node[below] {N};
      \path (b) node[left] {T};
      \path (c) node[above] {N};
      \path (d) node[above] {N};
      \path (e) node[right] {N};
      \path (f) node[right] {N};
      \path (g) node[right] {N};

      \draw (1.5,-.3) edge[snake=snake, segment amplitude=1.2pt] node[above]{\it etc.}(2.3,-.3);
      \draw (2.3,-.3) edge[->] (2.4,-.3);
    \end{tikzpicture}
  }
  \caption{\label{fig:sequence}A possible sequence of execution of the spanning forest algorithm.}
\end{figure}

\subsection{Correctness}
In this Section we establish some properties of the spanning forest algorithm, namely, that there is always exactly one root (token) in every tree, and no cycle can possibly occur.

\newpage
\begin{lemma}
  \label{lem:at-least-one}
  At any time, there is at least one token per tree.
\end{lemma}
\begin{proof}
  The lemma holds initially, when every node is the root of its own tree. Now observe that both {\em merging} and {\em circulation} operations perserve this property. Indeed, the application of $r_1$ merges two trees but suppresses one token, while $r_2$ just moves a token within the underlying tree. We can thus focus on the disappearance of edges. Whenever an edge $e$ disappears, either $e$ did not belong to a tree or it did. If it did not, nothing has to be done. If it did, then this tree is now split into two trees, one of which is left token-less. By rule $r_a$, whose application is immediate, a token is regenerated on the orphan side of that edge (edge state $1$). If several such edges had disappeared simultaneously, the same mechanism would have occurred relative to each fragment.\qed
\end{proof}

\begin{lemma}
  \label{lem:at-most-two}
  At any time, there is at most one token per tree.
\end{lemma}
\begin{proof}[By contradiction]
  The only rule leading to the creation of a token is $r_a$. Since the lemma holds initially, the presence of more than one token in a tree must result from one of these events:
  \begin{enumerate}
    \item Rule $r_a$ was applied despite the existence of another token in the tree.
    \item Rule $r_a$ was applied several times simultaneously in the tree.
  \end{enumerate}
  In the first case, the contradiction stems from the fact that $r_a$ is applied on the child endpoint of a lost edge. By construction, the token is thus on the other side and the local subtree is token free. In the second case, the contradiction is slightly less direct. Let $v$ and $v'$ be two nodes of a same tree, both of which have applied $r_a$ simultaneously. Three cases are possible regarding the relative position of $v$ and $v'$ in the tree:
  \begin{enumerate}
  \item
  \begin{enumerate}
  \item $v$ is an ancestor of $v'$. This is impossible because the application of $r_a$ by $v'$ results from the disappearance of its parent edge.
  \item $v'$ is an ancestor of $v$. Same argument for $v$.
  \item $v$ and $v'$ have a common ancestor. This is again impossible because the application of $r_a$ results from the disappearance of a parent edge, therefore neither $v$ nor $v'$ can have an ancestor at all.\qed
  \end{enumerate}
  \end{enumerate}
\end{proof}

\begin{theorem}
  At any time, there is exactly one token per tree.
\end{theorem}
\begin{proof}
  By Lemmas~\ref{lem:at-least-one} and~\ref{lem:at-most-two}.\qed
\end{proof}

\begin{theorem}
  At any time, the trees are cycle-free.
\end{theorem}
\begin{proof}
    The property holds initially. The only way an edge can be added to a tree is by means of applying $r_1$, which involves two tokens. By Lemma~\ref{lem:at-most-two}, there is at most one token per tree, thus at most one application of $r_1$ can occur at a time for a given tree, and the two tokens must belong to different trees. \qed
\end{proof}

\subsection{Discussion}
\label{sec:discussions}

The algorithmic principle introduced here is very general. 
In particular, the correctness of the properties we have considered so far does not depend on the order in which the edges are selected for interaction, nor whether some interactions should be favored over others (e.g. $r_1$ over $r_2$). 
\FloatBarrier
On the other hand, these aspects can have a tremendous impact on the ability of the trees to merge with each other and converge towards a single tree per connected component (remind that the network is expected to be partitioned in general).

\paragraph{Priority among rules:}
In general, given two neighbor nodes at a given time, there might be more than one eligible rule. This was not the case with this algorithm, since $r_1$ and $r_2$ have two incompatible guards (preconditions). However, the matter is worth being discussed. Priority among the rules could be understood in a {\em weak} sense, enforcing the fact that a rule should not be applied by {\em these two} nodes if they are able to apply another rule first. Another, much {\em stronger} sense of priority consists in forbidding a node to apply a given rule as long as another rule is applicable {\em with any} of its neighbors. 

Clearly, in the case of the spanning forest algorithm, merging should be preferred over circulation whenever possible. Enforcing strong priority would thus come to forbid the application of $r_2$ whenever $r_1$ can be applied. This behavior is expected to produce larger trees, but at the cost of a strong constraint on the scheduler (probing the state of an entire neighborhood prior to interaction). Without speculating on finer-grain implementations of our principle -- which is not the object of this paper -- we believe a strong priority mechanism remains somewhat natural in a wireless environment, where nodes routinely broadcast their state to all neighbors, in particular if we assume a synchronous communication model such as ${\cal LOCAL}$ or ${\cal CONGEST}$~\cite{peleg2000}. 

\paragraph{Role played by both nodes in an interaction:} The reader may have noticed that, in the definition of the circulation rule $r_2$, the guard of the rule is not tested on both sides. That is, $u$ implicitely plays the role of the left node, and $v$ that of the right node. As far as the present work is concerned, we do not want to impose a preferred way to solve this question, as it does not affect correctness. As a suggestion, the scheduler may select edges in a directed way (with a left node, and a right node), or the second direction systematically when an edge is selected and the rule is not applicable in the first direction.

\paragraph{High-level view of the process:} Assuming the token has equal probability to move to each neighbor (in the tree), we can regard the circulation as a random walk in the tree. Further, if we assume {\em strong} priority enforcement between $r_1$ and $r_2$, the circulation and merging processes turn into  a specific variant of coalescing random walks~\cite{cooper2012}. This point of view is the one we consider in the next section.

\section{Preliminary analysis\label{sec:analysis}}
In this section, we study the question of how frequent the mergings are. We only provide preliminary results and some thoughts about the complete analysis of this process (which is far beyond the scope of this paper). Hence, we characterize the number of token moves expected in a stationary regime, before a merging occurs between two given trees in a static context. This value is given as a function of their size and the number of edges connecting them (called {\em bridges}). \vspace{-6pt}

\subsection{Random walks in trees}
For the sake of analysis (and with loss of generality), we look at the process of merging and circulating tokens as a system of particles that perform random walks {\em in trees} and coalesce whenever they {\em meet}. Here, the concept of meeting between two particles is defined with a special meaning. Indeed, in most coalescing particle systems, two particles are said to meet if they happen to be located at a same node, whereas in our case, they meet if they are located at both endpoints of a same edge (remind that the tokens cannot travel beyond their trees).\vspace{-4pt}


\subsection{Bridges}
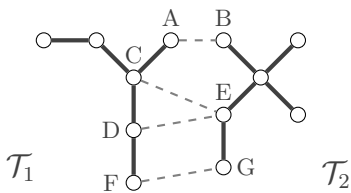
\begin{figure}[h]
  \centering
    \begin{tikzpicture}[-,>=stealth,auto,node distance=.7cm, scale=1]
      \tikzstyle{state}=[draw, circle, inner sep=2pt]
      \tikzstyle{lien}=[dashed, gray, thick]
      \tikzstyle{every node}=[]
      \node[state]         (L)                      {};
      \node[state]         (Z) [right of = L]       {};
      \node[state]         (C) [below right of = Z] {};
      \node[state]         (A) [above right of = C] {};
      \node[state]         (D) [below of = C]       {};
      \node[state]         (F) [below of = D]       {};
      \node[state]         (B) [right of = A]       {};
      \node[state]         (W) [below right of = B] {};
      \node[state]         (E) [below left of = W]  {};
      \node[state]         (K) [above right of = W] {};
      \node[state]         (I) [below right of = W] {};
      \node[state]         (G) [below of = E]       {};

      \tikzstyle{every path}=[ultra thick, black!75]
      \path (L) edge  node {} (Z)
            (Z) edge  node {} (C)
            (A) edge  node {} (C)
            (C) edge  node {} (D)
            (D) edge  node {} (F);
      \path (B) edge  node {} (W)
            (E) edge  node {} (W)
            (W) edge  node {} (K)
            (W) edge  node {} (I)
            (E) edge  node {} (G);
      \path (A) edge[lien]  (B)
            (D) edge[lien]  (E)
            (C) edge[lien]  (E)
            (F) edge[lien]  (G);
      \tikzstyle{every node}=[]
      \path (A)+(0,.3) node (labA) {A};
      \path (B)+(0,.3) node (labB) {B};
      \path (C)+(0,.3) node (labC) {C};
      \path (D)+(-.3,0) node (labD) {D};
      \path (E)+(0,.3) node (labE) {E};
      \path (F)+(-.3,0) node (labF) {F};
      \path (G)+(.3,0) node (labG) {G};
      \path (F)+(-1.5,.2) node (T1) {\large $\mathcal{T}_1$};
      \path (G)+(1.5,-.1) node (T2) {\large $\mathcal{T}_2$};
    \end{tikzpicture}\vspace{-6pt}
  \caption{Example of two trees sharing four bridges (dashed lines).\label{fig:spanningforest1}}
\end{figure}

Given two different trees $\mathcal{T}_1$ and $\mathcal{T}_2$, there may be some edges whose endpoints lie in $\mathcal{T}_1$ on one side, and $\mathcal{T}_2$ on the other side -- we call such edges {\em bridges}. Figure~\ref{fig:spanningforest1} shows an example of two trees that share four bridges.\vspace{-6pt}

As discussed in Paragraph~\ref{sec:discussions}, the enforcement of a {\em strong} notion of priority between merging and circulation allows one to assume that if two tokens are located on a same bridge, then merging occurs. (This is at least true in the case of two trees, which is the one addressed here.) Hence, the probability that merging occurs is that of having both tokens located at a same bridge. 

Let us denote by $Bridges(\mathcal{T}_1, \mathcal{T}_2)$ the set of edges $(u,v)$ such that $u \in E_{\mathcal{T}_1}$ and $v \in E_{\mathcal{T}_2}$. The probability that $\mathcal{T}_1$ and $\mathcal{T}_2$ merge at a given time is equal to:

\begin{equation}
  \label{eq:proba-sum}
  P_{merge(\mathcal{T}_1,\mathcal{T}_2)}=\sum_{(u,v) \in Bridges(\mathcal{T}_1, \mathcal{T}_2)} \footnotesize \scriptstyle P[\lambda(u)=T \wedge \lambda(v)=T].
\end{equation}

\subsection{Probability of being located at a node}
\label{ssec:random}
In a stationary regime, the probability for a token to be located at a given node $v$ in a graph $G$ (tree or not) is a well-known result in random walk theory, which only depends on the ratio between the degree of $v$, $d_G(v)$, and the sum of all degrees in $G$. 

In a tree $\mathcal{T}$, the probability a node $v$ hosts the token is thus
\begin{equation}
  \label{eq:proba-each}
 P(\lambda(v)=T) = \frac{d_{\mathcal{T}}(v)}{2|E_{\mathcal{T}}|} \vspace{-4pt}
\end{equation}
where $|E_{\cal T}|$ is the size of ${\cal T}$.
Keep in mind this value corresponds to the stationary regime (when the probabilities no more depend on the initial configuration). 




\subsection{Expected merging time in the stationary regime}
We are interested in the mean number of steps (token moves) required to merge the two trees, 
assuming the walks are in a stationary regime. Moreover, as trees are bipartite graphs, if both tokens move synchronously it may happen, depending on their initial position, that they never meet. Thus, we assume here that the moves are asynchronous (i.e., one at a time). Equations~\ref{eq:proba-sum} and~\ref{eq:proba-each} allow us to state that the probability for two trees $\mathcal{T}_1$ and $\mathcal{T}_2$ to merge at any step is
\begin{equation}
  \label{eq:proba-mixed}
  \footnotesize 
  P_{merge}(\mathcal{T}_1,\mathcal{T}_2)=\sum_{\{(u,v) \in Bridges(\mathcal{T}_1,\mathcal{T}_2)\}} \frac{d_{\mathcal{T}_1}(u)}{2|E_{\mathcal{T}_1}|} \times \frac{d_{\mathcal{T}_2}(v)}{2|E_{\mathcal{T}_2}|}
\end{equation} 

\noindent which in turn gives the \emph{expected merging time} in number of steps, \linebreak as  $E_{merge}(\mathcal{T}_1,\mathcal{T}_2)=P_{merge}(\mathcal{T}_1,\mathcal{T}_2)^{-1}$.\smallskip


However limited, a quick look at these results teaches us some preliminary facts. First, the merging time of two trees of size $n$ in which the nodes degrees $d$ are fairly distributed is in $O(\frac{n^2}{nbBridges\cdot d^2})$. The $d^2$ term could actually be omitted if we consider that degrees are bounded by some constant (a fair assumption in most wireless networks). Whence a time of $O(\frac{n^2}{nbBridges})$ steps. Whether this time is linear or quadratic in the sizes of the trees depends on the number of bridges (e.g. merging time is linear if the number of bridges is in $O(n)$; it is quadratic if that number is constant; etc.). That in turn, depends on the networking scenario which is considered, and in particular, what {\em mobility model} is used.

A deeper look is required to understand the behavior of this process. We expect it to be quite difficult to analyze in the general case. Not only the algorithm involves much more than two trees in general, but it is intended to run over highly dynamic topologies, where splittings and mergings occur concurrently. In fact, the metric of interest might be different than convergence time, since the merging process is never expected to converge. A better metric here could be the average number of trees per connected component in a stationary regime.

\section{Conclusion\label{sec:conclusion}}

This paper proposed a new mechanism for building and maintaining a forest of spanning trees in highly dynamic networks. The originality of the approach is that the construction is a perpetual ongoing process that takes place at the same time as the trees are used. The principle is very general and relies on token circulation techniques that turns \emph{splittings} and \emph{mergings} of the trees into purely localized phenomenons. After presenting the algorithm using a coarse grain interaction model, we provided some preliminary observations on the analysis of the corresponding process, regarded for the occasion as a system of coalescing random walks. A deeper analysis of this process is still far from reach, and we expect it to be technically challenging in the general case. As the process is never expected to converge, completion time is not the most relevant metric here (however its characterization in a static and connected context might already be very insightful). Characterizing the average number of trees per connected component in the stationary regime seems to be the relevant metric.

Besides analysis, an avenue of research is to transpose the algorithm into finer-grain communication models. We believe this can be done, at least in synchronous message passing models. Finally, de-randomizing the way tokens circulate (e.g. using Propp machine-like mechanisms) may lower the cover time, and possibly, speed-up the merging process. These questions are open.




\end{document}